\documentclass[aps,prb,reprint,superscriptaddress]{revtex4-1}

\usepackage{graphicx}
\usepackage{epsfig}
\usepackage{amsmath,amsthm}
\usepackage{textcomp}
\usepackage{float}
\usepackage{color}
\usepackage{empheq}
\usepackage{multirow}

\begin{document}

\title{Cu-Sb dumbbell arrangement in the spin-orbital liquid candidate Ba$_3$CuSb$_2$O$_9$}
\author{Michaela Altmeyer}
\affiliation{Institut f\"ur Theoretische Physik, Goethe-Universit\"at Frankfurt, Germany}

\author{Frederic Mila}
\affiliation{Institute of Physics, Ecole Polytechnique F\'ed\'erale de Lausanne (EPFL), CH-1015 Lausanne, Switzerland}

\author{Andrew Smerald}
\affiliation{Institute of Physics, Ecole Polytechnique F\'ed\'erale de Lausanne (EPFL), CH-1015 Lausanne, Switzerland}

\author{Roser Valent\'i}
\affiliation{Institut f\"ur Theoretische Physik, Goethe-Universit\"at Frankfurt, Germany}

\date{\today}

\begin{abstract} 

The absence of both spin freezing and of a static Jahn-Teller effect have led to the proposition that Ba$_3$CuSb$_2$O$_9$ is a quantum spin-orbital liquid.
However, theoretical understanding of the microscopic origin of this behavior has been hampered by a lack of consensus on the lattice structure.
Cu ions have been proposed to realize either a triangular lattice, a short-range ordered honeycomb lattice or
 a disordered lattice with stripelike correlations.
Here we analyze the stability of idealized versions of these
arrangements using density functional theory.
We find stripe order of Cu ions
to be energetically favoured, hinting towards stripelike local Cu-Cu arrangements, while long-range order is presumably hindered due to disorder effects.
Furthermore, we find evidence of significant interlayer interactions between Cu-Sb dumbbells, which affects the out-of-plane arrangement. 
Analysis of the relaxed crystal structures, electronic properties and
tight-binding parameters provides clues as to the nature of the Jahn-Teller distortions.
\end{abstract}
\maketitle

\section{Introduction} 


In recent years there has been an enormous amount of effort devoted to the search
for materials with quantum spin liquid phases~\cite{Lacroix,Mendels,Kanoda,Khaliullin,Modic,Takagi,Coldea,Winter}.
A special class of quantum liquids are those where not only the spins but also the orbitals remain disordered
and fluctuating at low temperatures, resulting in a spin-orbital
liquid state~\cite{Ishihara1997,Khaliullin2000,Khomskii2003,Corboz2012,Silverstein2014}.

Recently,  Ba$_3$CuSb$_2$O$_9$ has been suggested as a promising candidate to realize
 spin-orbital liquid
behavior~\cite{Zhou2011,Nakatsuji2012,Quilliam2012,Ishiguro2013,Katayama2015,Han2015,
Wakabayashi2016,Li2016,Sugii2016}.
Thermodynamic measurements~\cite{Zhou2011,Nakatsuji2012}, muon spin rotation ($\mu$SR)~\cite{Quilliam2012}, and neutron-scattering experiments~\cite{Nakatsuji2012}  find no  spin ordering
or spin freezing down to temperatures as low as 20mK.
Electron spin resonance (ESR) and x-ray diffraction measurements find no evidence of
 a cooperative Jahn-Teller transition, implying that, not only  spins, but also orbitals fail to 
order~\cite{Nakatsuji2012}.
Furthermore, variable frequency ESR measurements suggest that the Jahn-Teller distortions remain dynamic down to low temperatures, rather than simply freezing and forming a glass-like arrangement~\cite{Han2015} .
This implies that the orbital degrees of freedom fluctuate at low temperature, and the ESR measurement of the timescale of this fluctuation is consistent with extended x-ray-absorption fine-structure (EXAFS) and x-ray measurements~\cite{Nakatsuji2012}.


\begin{figure}
\centering
\includegraphics[width=0.4\textwidth]{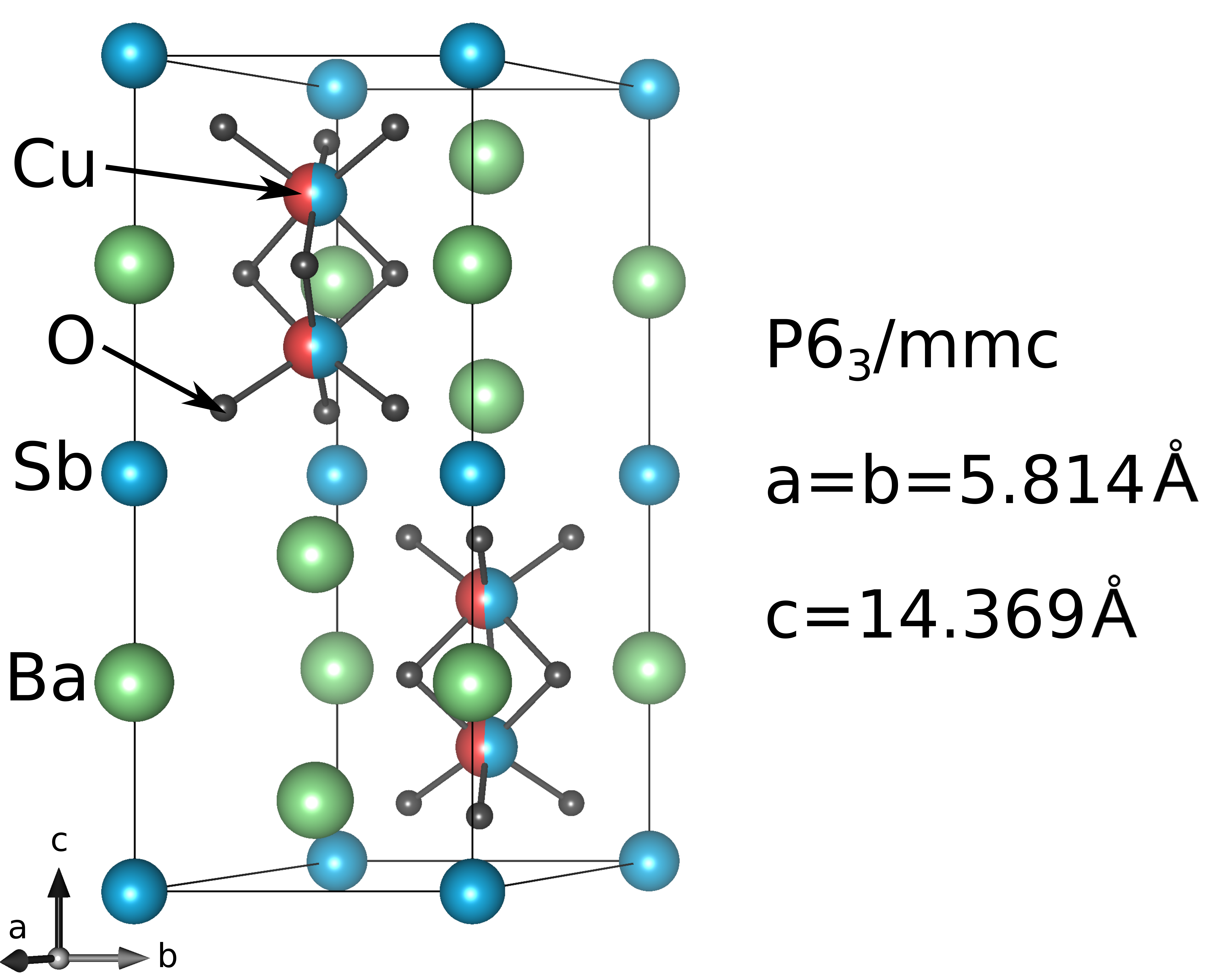}
\caption{\footnotesize{
The crystal structure of Ba$_3$CuSb$_2$O$_9$, as revealed by x-ray- and neutron-diffraction experiments~\cite{Nakatsuji2012}.
Cu-Sb dumbbells (red/blue) form a triangular lattice, but each dumbbell is in one of two possible orientations, with either the Cu or Sb ion on top.
Experimentally, in the absence of long-range order, it is difficult to determine the local arrangement of the dumbbells.
}}
\label{fig:ExpStr}
\end{figure}

While the spin and orbital degrees of freedom in Ba$_3$CuSb$_2$O$_9$
are primarily associated with the Cu ions,
 the important subunits of the system are  Cu$^{2+}$-Sb$^{5+}$ charged dumbbells
surrounded by  oxygen bioctahedra (see Fig.~\ref{fig:ExpStr}).  These
dumbbells form a triangular lattice and  two orientations of the dumbbell
are possible, with either  Cu or  Sb  on top.
Despite intensive experimental and theoretical investigations,
 no clear consensus has been reached so far on the arrangement of these dumbbells. Such a knowledge
is necessary in order to unveil the microscopic origin of the
 spin-orbital liquid behavior~\cite{Nasu2013,Smerald2014,Smerald2015}.
There have been three main proposals:
(i) the dumbbells  have all the same orientation and the Cu ions thus form a triangular lattice~\cite{Koehl1978,Zhou2011};
(ii) the dumbbells are disordered but form a honeycomb structure at short
lengthscales~\cite{Nakatsuji2012};
(iii) a stripe structure of Cu spins is energetically favorable
with the dumbbells freezing in a disordered state with short-range stripelike correlations~\cite{Smerald2015}.


 In this work we compare these different options 
via density functional theory (DFT) calculations of
 idealized, long-range-ordered dumbbell arrangements
with different three-dimensional stackings
and show that the stripe configuration (iii) has the lowest energy. 
Even though in   Ba$_3$CuSb$_2$O$_9$ 
 the dumbbells are disordered, our results hint at stripelike arrangements at the local level~\cite{Smerald2015}.
Finally, starting from the idealized structure with long-range dumbbell order, we discuss both the nature of the expected Jahn-Teller distortion and the electronic properties.

\section{Crystal structures}

A number of different arrangements of  Cu$^{2+}$-Sb$^{5+}$ dumbbells have been proposed in the
past and it is useful to summarize the evidence for each of these.

Historically, Ba$_3$CuSb$_2$O$_9$ was proposed  to crystallize
 in the centrosymmetric space group $P6_3mc$, and to have a parallel alignment of the 
dumbbells~\cite{Koehl1978,Zhou2011},
so that the Cu ions form a triangular lattice.
However, this arrangement would be  associated with a large electrical dipole
moment and is not consistent with pyroelectric
measurementa~\cite{Nakatsuji2012}.
Furthermore, previous DFT calculations showed
 that this arrangement was  energetically unfavorable with respect
to honeycomb dumbbell arrangements~\cite{Shanavas2014}.

Additional work on the refinement of the crystal structure suggested that the space group is 
 the centrosymmetric $P6_3/mmc$~\cite{Nakatsuji2012}.
The appearance of partial occupancies of Cu/Sb positions
 implies that there is no long-range ordering of the dumbbells,
but one expects an equal number of dumbbells in each orientation so as to
be compatible with the absence of internal dipole moments observed in
pyroelectic measurements.
 X-ray diffraction data show no Bragg peaks associated with the dumbbell ordering, confirming that the dumbbells are disordered~\cite{Nakatsuji2012}.
Diffuse x-ray scattering measurements suggest short-range honeycomb order of the dumbbells, in which the central dumbbell of a hexagonal plaquette is oriented anti-parallel to the other dumbbells~\cite{Nakatsuji2012}.
However, it should be noted that the approximate domain size was found to be $\sim 10$ \AA, which is less than twice the spacing between neighboring dumbbells.
The proposal of honeycomb short-range order inspired a number of theoretical studies of the spin and orbital physics, but it was found that long-range orbital order is difficult to avoid~\cite{Nasu2013,Smerald2014}.

Finally, a recent theoretical study proposed that the lowest-energy dumbbell
arrangement consists of stripes, with a parallel dumbbell alignment within each
stripe and antiparallel alignment between neigboring
stripes~\cite{Smerald2015}.
This suggestion was based on solving a highly simplified two-dimensional (2D) model, in which dumbbells interact only via their charge degrees of freedom.
It was further discussed that the materials do not reach this ground-state configuration, but instead freeze into a disordered configuration with short-range, stripelike correlations, dubbed a ``branch'' lattice.
This picture was found to be consistent with the x-ray-diffraction data reported in Ref.~[\onlinecite{Nakatsuji2012}], which was initially taken as evidence of a honeycomb structure, as well as
an  x-ray study of the pair distribution function~\cite{Wakabayashi2016}.
In fact this ``branch'' lattice is naturally tuned to a percolation critical point and preliminary calculations suggest that this type of correlated disorder could drive the formation of a spin-orbital liquid~\cite{Smerald2015}.
There is a clear need for a first-principles calculation to evaluate whether such a stripe-ordered configuration is indeed lower in energy than other competing structures.

\section{Methods}

We perform DFT calculations
within the PBE generalized gradient approximation (GGA), the spin-polarized GGA and
GGA+$U$, starting from a set of idealized crystal structures with different dumbbell ordering patterns. 
In order to exclude the possibility that the correlations might yield different effects when more sophisticated functionals are considered, we have performed test calculations using the HSE~\cite{HSE} hybrid functional, which we found to perfectly follow the trends of our spin-polarized GGA and GGA+$U$ calculations, while the stability of the stripy alignments is even enhanced. 
For each structure we use a high energy cut-off of 520~eV for the plane waves and relax the internal coordinates~\cite{comment2} using the projector augmented wave method, as implemented in the Vienna Ab initio Simulation Package (VASP~\cite{VASP}). 
As the lattice parameters have been confirmed in several measurements~\cite{Koehl1978,Zhou2011,Nakatsuji2012} we keep them fixed to their experimental values. 
We have verified that our results on structure stability and Jahn-Teller distortions remain valid also in the case of a full relaxation, where additionally the volume and shape of the unit cell are relaxed~\cite{Shanavas2014}.
The total energies were checked using the all-electron full-potential localized orbital code (FPLO~\cite{FPLO}), which was also employed to calculate the electronic band structures and density of states.
In previous DFT studies correlation effects were found to play an important role, not only to achieve insulating behavior, but also to allow for the formation of distinct Jahn-Teller distortions of the oxygen octahedra~\cite{Shanavas2014}.
Here we compare our GGA calculations to Coulomb corrected GGA+$U$ simulations,
 where the correlations on the Cu atoms are explicitly included via a Hubbard term.\cite{Dudarev1998} 
We choose  $U=4$~eV  in accordance with previous studies.\cite{Shanavas2014}
The total energies are converged on a $6\times 6 \times 6$ and $4 \times 4\times 4$ k mesh for the stripy (four formula units per unit cell) and honeycomb (six formula units per unit cell) arrangement of the dumbbells, respectively.

\section{Comparison of energies}

%
\begin{figure}
\centering
\includegraphics[width=\columnwidth]{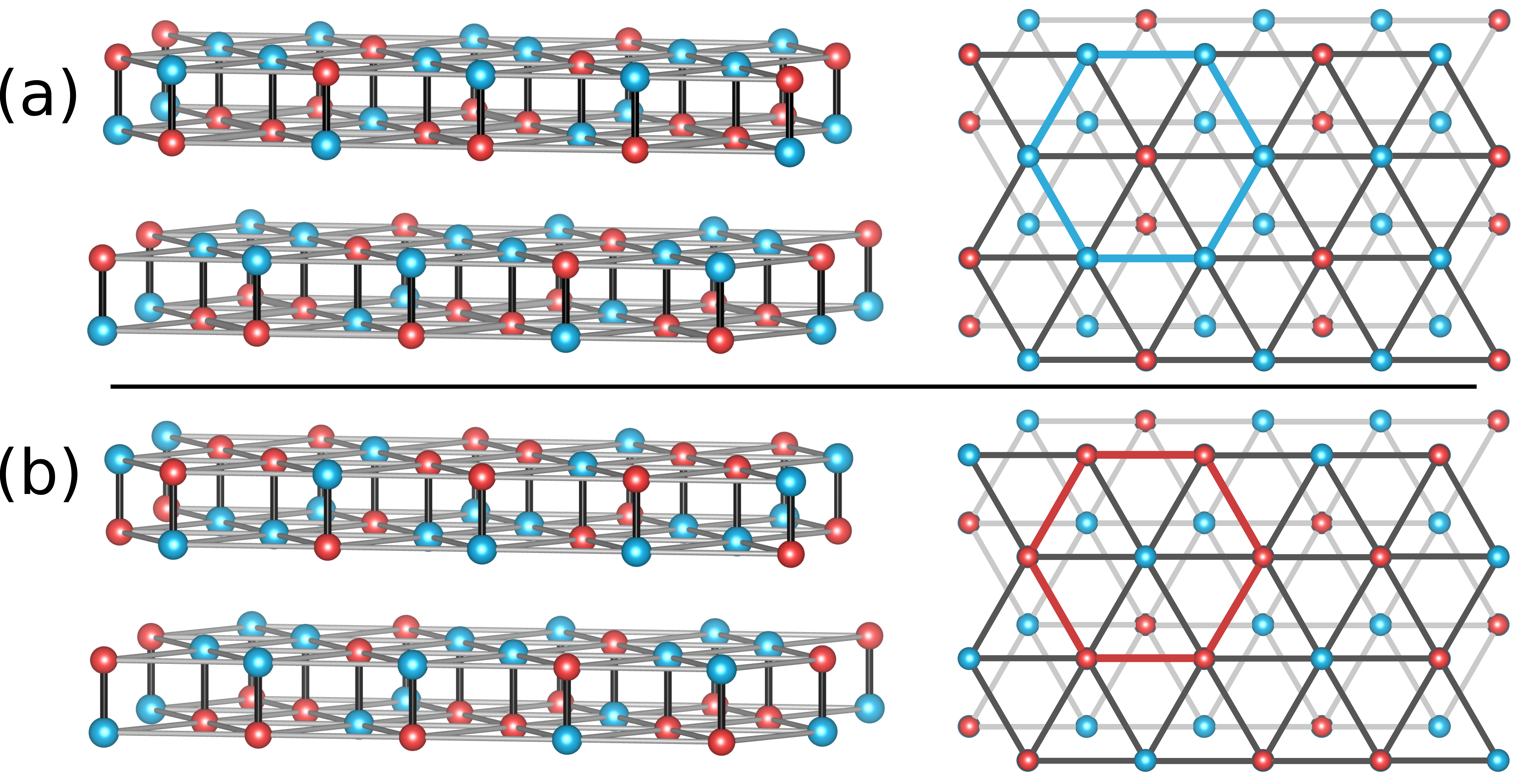}
\caption{\footnotesize{
The two honeycomb arrangements of Cu-Sb dumbbells taken as starting points for the density functional theory calculations.
Cu ions are shown in red and Sb in blue.
(a) Side and top views of the {\it honey-a} structure, which has a net dipole moment.
(b) Side and top views of the {\it honey-b} structure, where the upper plane has been flipped with respect to the {\it honey-a} structure.
In the {\it honey-b} structure the net dipole moment cancels between neighboring planes.
}}
\label{fig:honey}
\end{figure}
\begin{figure}
\centering
\includegraphics[width=\columnwidth]{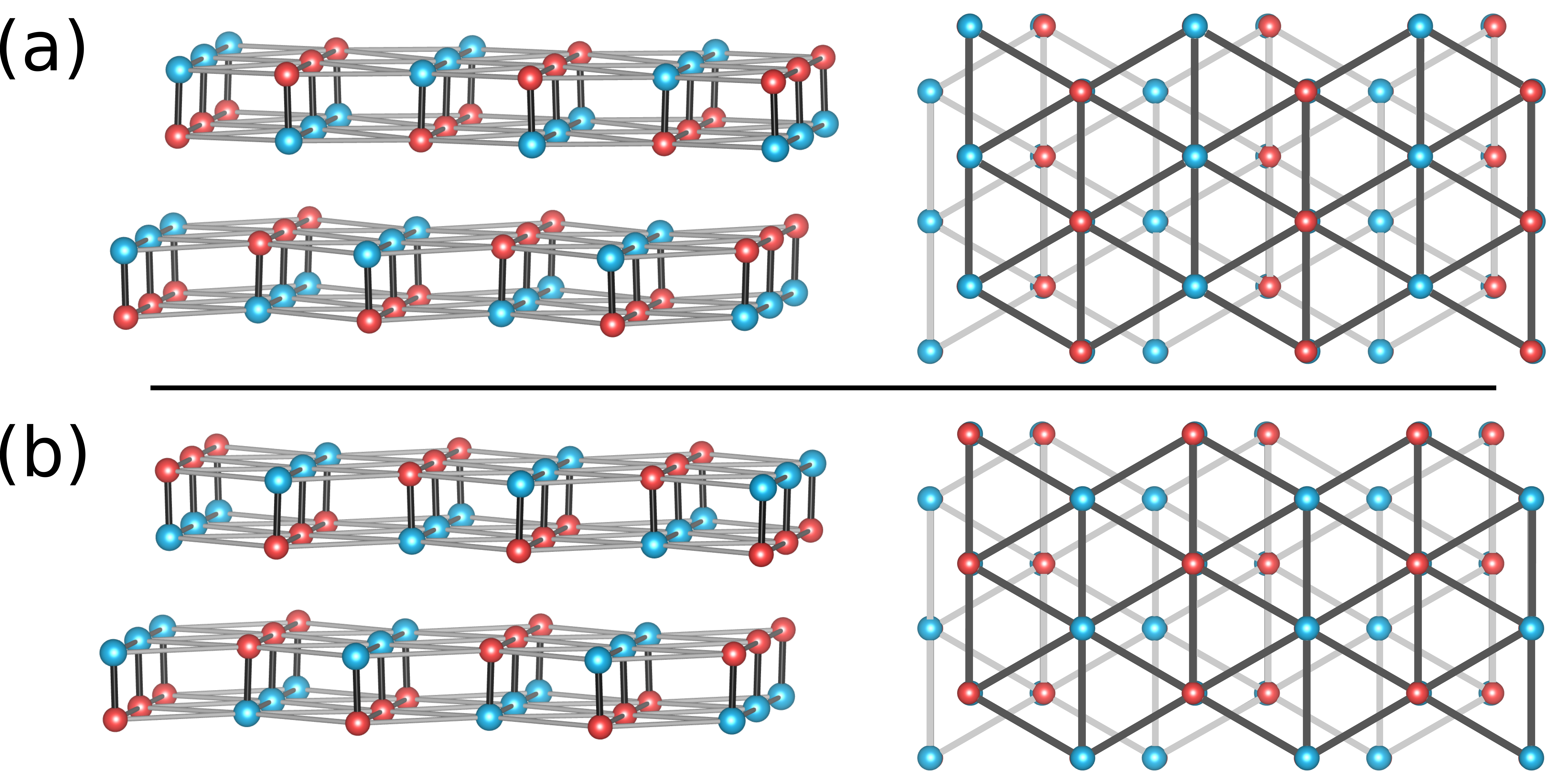}
\caption{\footnotesize{
The two stripe arrangements of Cu-Sb dumbbells taken as starting points for the density functional theory calculations.
Cu ions are shown in red and Sb in blue.
(a) Side and top views of the {\it stripy-a} structure.
This configuration is chosen so as to minimize the energy of the electrostatic interaction between neighboring planes.
(b) Side and top views of the {\it stripy-b} structure, where the upper plane has been flipped with respect to the {\it stripy-a} structure.
}}
\label{fig:stripy}
\end{figure}

In order to discern which of the proposed dumbbell arrangements is lowest in energy,
  we consider four different structures with long-range dumbbell order, shown in Figs.~\ref{fig:honey} and \ref{fig:stripy}, relax their internal coordinates, and compare their energies.
Although in Ba$_3$CuSb$_2$O$_9$  the dumbbells are not long-range ordered, the idea is that energy differences between these idealized structures will give an indication as to the type
of local correlations in the system.

The first structure we consider, {\it honey-a}, has a ferrimagnetic
honeycomb arrangement of the dumbbells in each layer,
in the sense that there is a 2:1 ratio between the two dumbbell 
orientations (Fig.~\ref{fig:honey}(a)).
The neighboring layers are chosen to have the same ferrimagnetic ordering
with a resulting  net dipole moment associated with the configuration.
This arrangement was considered in Ref.~[\onlinecite{Shanavas2014}] and it was found to be energetically favorable over the parallel alignment of all dumbbells.

The second structure, {\it honey-b}, has the same honeycomb arrangement in the layers
as {\it honey-a}  but the stacking of layers is such that there is no net dipole moment 
(Fig.~\ref{fig:honey}(b)).
From a naive electrostatic picture this would appear to be higher in energy than {\it honey-a}.
However, this consideration doesn't take into account the exchange energy, and,
as noted in Ref.~[\onlinecite{Nakatsuji2012}], interlayer exchange pathways
between Cu ions at the top of the lower layer and bottom of the upper layer are
comparable with those within the layers, and thus could be important. 

The third structure,  {\it stripy-a}, has a stripe arrangement of the dumbbells
in each layer (Fig.~\ref{fig:stripy}(a)).
The interlayer stacking is chosen such as to minimize the electrostatic energy between neighboring layers.

The fourth structure,  {\it stripy-b}, also has a stripe
 arrangement of the dumbbells in each layer (Fig.~\ref{fig:stripy}(b)).
However, a different interlayer stacking is chosen, which is the only period-2 alternative to {\it stripy-a}, assuming that stripes in neighboring layers run parallel to one another.

\begin{figure}
\includegraphics[width=\columnwidth]{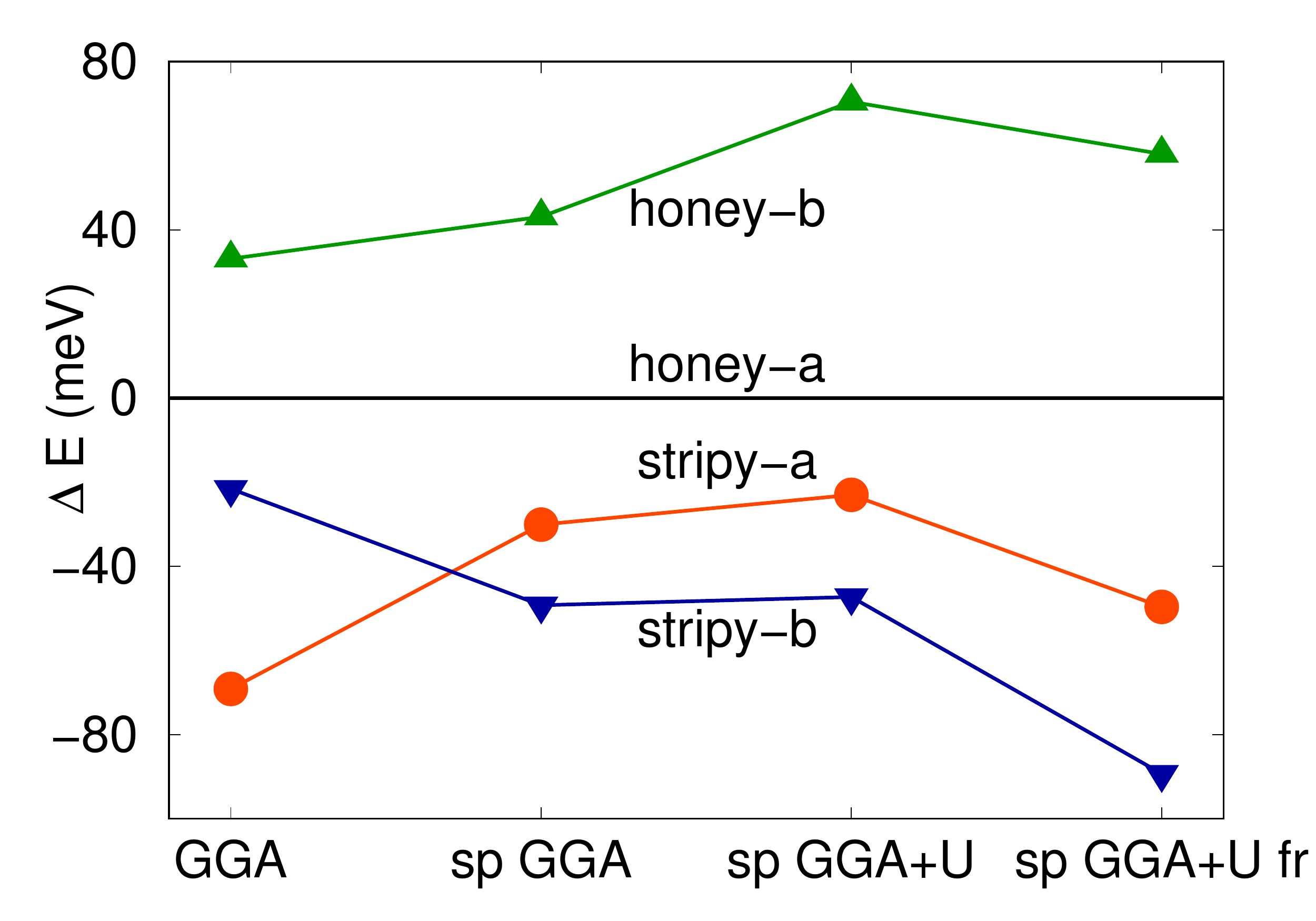}
\caption{\footnotesize{
Total energy per formula unit of different dumbbell configurations, measured relative to that of the {\it honey-a} structure (see Fig.~\ref{fig:honey}(a)).
Three different functionals are used, non spin-polarized GGA, spin-polarized GGA and GGA+$U$ in a ferromagnetic setup, and these give broadly consistent results. Additionally, we show
the results for the full relaxation (fr), which is even enhancing the observed energy splittings.
The lowest energy configuration is found to be the {\it stripy-b} structure (see Fig.~\ref{fig:stripy}(b)).
}}
\label{fig:toten}
\end{figure}

Comparing the total energies of
the two honeycomb structures, we find that {\it honey-a} has a considerably lower energy than {\it honey-b} of at least $ 30$~meV per formula unit (see Fig.~\ref{fig:toten}).
This is in agreement with naive electrostatic considerations, and shows the importance of the interlayer effects on the dumbbell structure.

The {\it stripy-a} and {\it stripy-b} structures are, interestingly, 
 considerably lower in energy than {\it honey-a} by at least $20$~meV per formula unit (see Fig.~\ref{fig:toten}).
When using pure GGA, the {\it stripy-a} structure is lower in energy than {\it stripy-b}, as expected from a simple electrostatic picture.
However, spin-polarized GGA and spin-polarized GGA+$U$ show that exchange effects favour the {\it stripy-b} structure, and the energy is better by approximately $20$~meV per formula unit.

Thus,  {\it stripy-b} has the lowest energy, which is  suggestive that local correlations will be of this type in the material.

\section{Jahn-Teller effect}

\begin{figure}
\centering
\includegraphics[width=0.95\columnwidth]{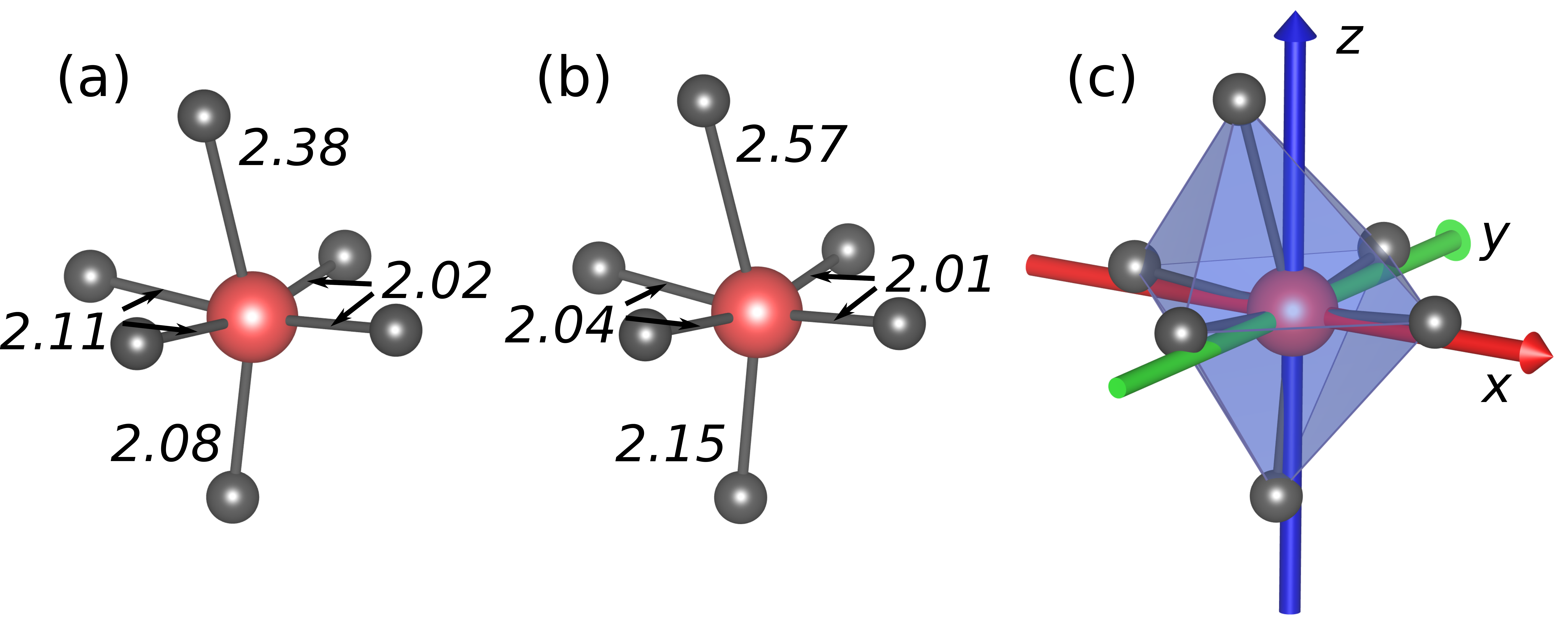}
\caption{\footnotesize{
Local arrangement of the oxygen (grey) octahedra surrounding a Cu ion (red). The Jahn Teller distortion as obtained in the spin-polarized GGA relaxation of the
(a) {\it stripy-a}  and 
(b)  {\it stripy-b} structure (see Fig.~\ref{fig:stripy}). 
The JT distortion is more pronounced in the {\it stripy-b} structure with 1 long bond of 2.57$\AA$, and 5 shorter bonds of between $2.01-2.15 \AA$.
(c) The local coordinate system used to derive the orbital contribution to the bands.
}}
\label{fig: JTdistloccoo}
\end{figure}

One of the most interesting, and most discussed, aspects of Ba$_3$CuSb$_2$O$_9$ is the nature of the Jahn-Teller effect.
While DFT calculations cannot simulate the dynamic effect that may be present in the materials\cite{Han2015}, it is useful to understand the nature and robustness of the static Jahn-Teller predicted by DFT for the idealized structures

In the {\it stripy-b} structure, we find that there is a considerable static Jahn-Teller distortion, and this is independent of the choice of functional.
One of the six Cu-O bonds is considerably longer than the other five, and the bond lengths are shown in Fig.~\ref{fig: JTdistloccoo}(b).
The bond opposed to the long bond is found to be slightly larger than the bonds lying in the perpendicular plane.
In the {\it stripy-a} structure, the Jahn-Teller distortion is similar, but a little less pronounced (see Fig.~\ref{fig: JTdistloccoo}(a)).

%

\section{Electronic properties}

\begin{figure}[t!]
\centering
\includegraphics[width=\columnwidth]{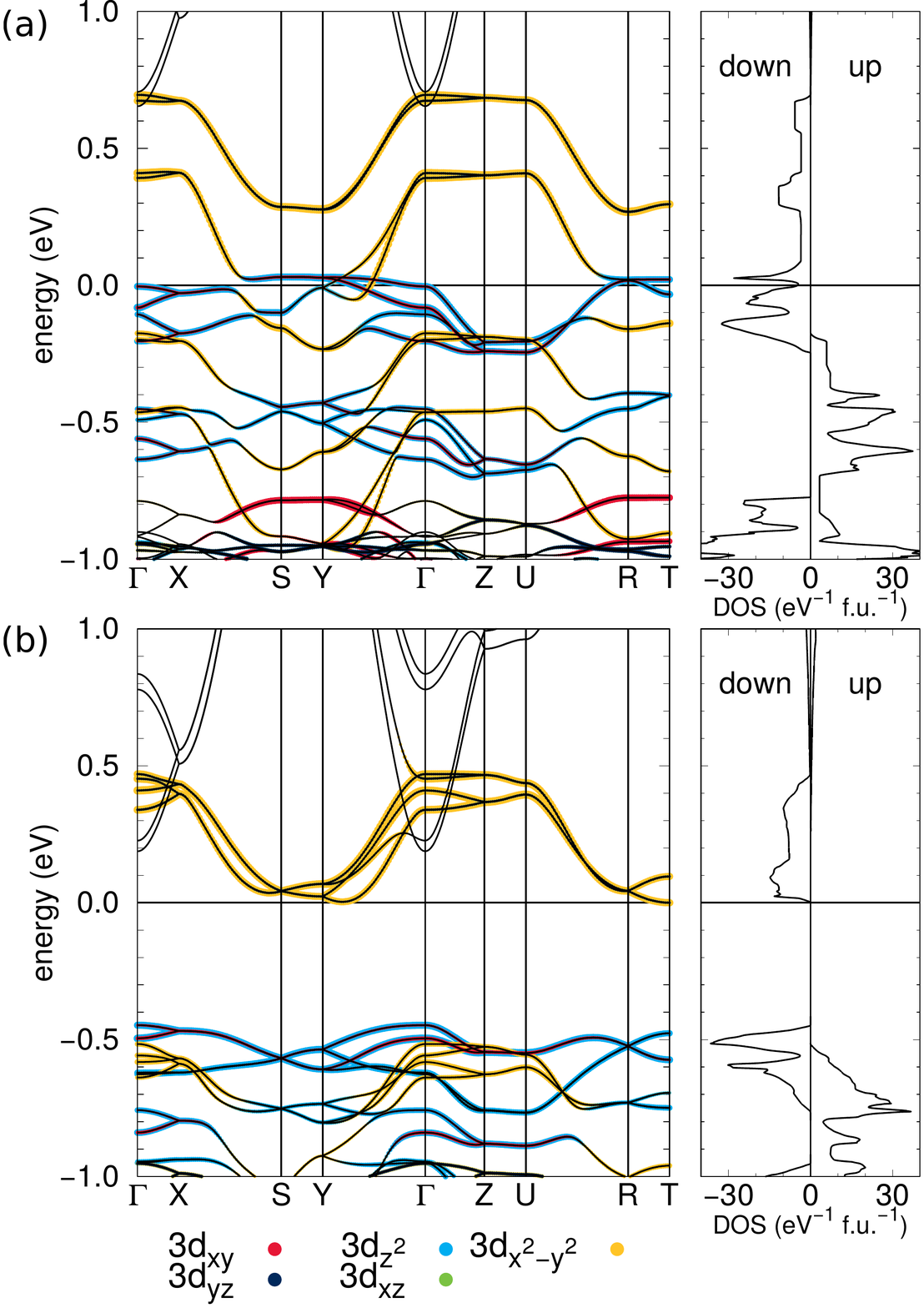}
\caption{\footnotesize{
Band structure and density of states of Ba$_3$CuSb$_2$O$_9$, calculated using spin-polarized GGA.
(a) Band structure associated with {\it stripy-a} dumbbell ordering (see Fig.~\ref{fig:stripy}(a)).
(b) Band structure associated with {\it stripy-b} dumbbell ordering (see Fig.~\ref{fig:stripy}(b)).
The colour scheme assigns a predominant character to the bands associated with the Cu d electrons.
The density of states is shown on the right-hand side.
}}
\label{fig:bands_stripynew}
\end{figure}

DFT calculations give access not only to the nature of the crystal structure, but also to the electronic band structure.

Common electronic properties of all
 considered crystal structures are:
 (i) a strong $e_g$-$t_{2g}$ splitting of the Cu $3d$ orbitals;  (ii) a Jahn-Teller distortion
 which induces splitting of the partially filled $e_g$ bands; (iii) the bands near the Fermi level
are predominantly of $d_{x^2-y^2}$ and $d_{z^2}$ character, but there are also small contributions
 from the oxygen $p$ orbitals and, due to additional distortions of the octahedra, 
 from Cu $t_{2g}$ orbitals.

The band structure associated with the {\it stripy-a} structure is shown in Fig.~\ref{fig:bands_stripynew}(a).
At the level of spin-polarized GGA, there is no gap.
In such a case it is not unreasonable that additional correlation effects could drive a dynamic Jahn-Teller distortion.

The band structure of the {\it stripy-b} structure is shown in
Fig.~\ref{fig:bands_stripynew}(b), and shows considerable differences from the
case  of the {\it stripy-a} structure, demonstrating the importance of
the interlayer dumbbell arrangement.
In particular,  there is a gap of about $0.5$~eV between the $e_g$ bands already
at the level of spin-polarized GGA calculations.
Using the local coordinate system shown in Fig.~\ref{fig: JTdistloccoo}(c), we 
observe that the band with predominantly $d_{z^2}$ character is fully occupied, while the $d_{x^2-y^2}$ minority-spin band is empty.

Finally, we determine the tight-binding parameters associated with the different crystal structures, via the Wannier functions~\cite{FPLO} .
In both the {\it stripy-a} and {\it stripy-b} structures, the dominant intralayer hopping path for the single hole in the $e_g$ band is along the stripe direction, and mediated by the $d_{x^2-y^2}$ orbitals.
Hopping between neighboring stripes within the same layer is highly suppressed, and this is in agreement with a simple analysis of exchange pathways\cite{Nakatsuji2012,Smerald2014}.
The main difference between the {\it stripy-a} and {\it stripy-b} structures is unsurprisingly in the interlayer hopping.
In the {\it stripy-a} the lobes of the $d_{x^2-y^2}$ orbitals in neighboring layers point towards each other, yielding hopping integrals as high as $t\gtrsim150$~meV.
In the {\it stripy-b} structure this is not the case, and the interlayer hopping is highly suppressed relative to the {\it stripy-a} case.
%

\section{Discussion}

The results of the DFT calculations suggest that the lowest-energy dumbbell configuration is the {\it stripy-b} structure.
This of course comes with the caveat that our search of possible dumbbell structures has not been exhaustive, but we have checked the most likely candidates proposed in the literature.
Also, while DFT calculations are not able to take all correlation effects into account, the trend we find is that the more carefully the correlation is treated, the larger the energy difference between the stripy and honeycomb arrangements.

Since all reported materials have dumbbell disorder, it is necessary to explain why this is the case, and what can be learned about the disordered structure from knowledge of the lowest-energy structure.
In Ref.~[\onlinecite{Smerald2015}] it was proposed that the dumbbells are flippable close to the synthesis temperature and then freeze at lower temperature into a disordered configuration.
A useful way to think about this is in terms of an effective Ising interaction between the dumbbells, which are constrained to have either Cu$^{2+}$ or Sb$^{5+}$ on top.
Knowledge of the lowest-energy dumbbell configuration is equivalent to knowing the ground state of the effective Ising model, and this provides information about the nature of the disordered phase above the ordering transition.
This of course relies on the assumption that the effective Ising interaction does not vary wildly with temperature, which seems reasonable given the absence of a structural transition\cite{Nakatsuji2012}.

Triangular-lattice Ising models with a stripe ground state are guaranteed by symmetry to have a first-order phase transition~\cite{Korshunov2005}, and it was shown in Ref.~[\onlinecite{Smerald2015}] that the disordered state that can be expected above this transition remembers the stripe nature of the ground state.
By this we mean that there are very few hexagonal plaquettes of six aligned dumbbells, and Cu ions therefore form a ``branch'' lattice~\cite{Smerald2015}.
This picture is consistent with Monte Carlo reconstruction of the experimentally determined x-ray pair-distribution function~\cite{Wakabayashi2016}.

The first-order transition into the stripe dumbbell structure can be contrasted with the probable second-order transition into the honeycomb-dumbbell structure\cite{Nienhuis1984}.
The first-order nature, and in particular the approximate topological constraint related to it\cite{Smerald2016}, will likely make it impossible to reach the stripe-ordered dumbbell ground state, no matter how slowly the crystal is cooled.
This is in contrast with the honeycomb dumbbell ground state, which would presumably be relatively easy to reach by reducing the cooling rate, due to the continuous nature of the transition.
This helps to explain why it has been so hard experimentally to find an ordered dumbbell structure.

Having discussed the likely structure of the disordered dumbbells, it is interesting to consider the nature of the spin-orbital state that forms at low temperature, where the dumbbells are frozen. 
Of particular interest has been the experimental observation that there is no long-range ordering of the Jahn-Teller distortions~\cite{Nakatsuji2012,Katayama2015}, and that rather than forming an orbital glass, the oxygen octahedra appear to remain fluctuating at low temperature with a dynamic timescale of approximately 100ps~\cite{Han2015}.
This is in contrast to our finding for the {\it stripy-b} structure, which shows a static Jahn-Teller distortion, with a sizable gap at the Fermi energy that is unlikely to be overcome by additional correlation effects (see Fig.~\ref{fig:bands_stripynew}).
One conclusion that can be drawn from this is that disorder is likely to be a key component of the orbitally fluctuating state, and therefore must be built into theoretical models from the start.

\section{Summary}

Using DFT we have explored how Cu-Sb dumbbells are arranged in the quantum spin-orbital liquid candidate Ba$_3$CuSb$_2$O$_9$.
This involved a comparison between different idealized dumbbell arrangements, and it was found that an in-plane stripe arrangement is considerably lower in energy than other candidate structures.
Experimentally realised crystals have dumbbell disorder, and our DFT calculations point to the presence of short-range stripelike correlations.

The main reason for studying the dumbbell structure is that it is very important for understanding the spin and orbital fluctuations that have been experimentally observed\cite{Han2015}.
The finding that the lowest energy {\it stripy-b} structure has a static Jahn-Teller distortion and a large gap of approximately $0.5$~eV at the Fermi level, suggests that the idealized structure does not form a spin-orbital liquid.
This in turn points towards the importance of disorder to explain the observed dynamic Jahn-Teller effect\cite{Smerald2015}.

By clarifying the arrangement of the Cu ions in Ba$_3$CuSb$_2$O$_9$ we hope to provide a good starting point for theories that aim to understand the nature of the experimentally observed spin and orbital fluctuations.


\begin{acknowledgments}
We thank L. Balents, C. Broholm and S. Nakasutji for useful discussions.
MA and RV thank the German Research Foundation (Deutsche Forschungsgemeinschaft) for support through grant SFB/TR49. 
AS and FM thank the Swiss National Science Foundation and its SINERGIA network ``Mott physics beyond the Heisenberg model'' for financial support.
MA, FM and RV further acknowledge partial support by the Kavli Institute for Theoretical Physics at the University of California, Santa Barbara under National Science Foundation Grant No. PHY11-25915.

\end{acknowledgments}





\begin{thebibliography}{99}
\bibitem{comment2} In contrast to the calculations by Shanavas et al., where additionally the lattice parameters and the shape of the unit cell are optimized.
\end{thebibliography}
\end{document}